\documentclass[journal]{IEEEtran}
\IEEEoverridecommandlockouts
\usepackage[noadjust]{cite}
\usepackage{amsmath,amssymb,amsfonts,amsthm}
\usepackage{algorithm}
\usepackage{algpseudocode}
\usepackage{graphicx}
\usepackage{textcomp}
\usepackage{xcolor}
\usepackage{bm}
\usepackage{hyperref}
\usepackage{siunitx}
\usepackage[normalem]{ulem}
\usepackage{accents}
\usepackage{cleveref}

\def\BibTeX{{\rm B\kern-.05em{\sc i\kern-.025em b}\kern-.08em
    T\kern-.1667em\lower.7ex\hbox{E}\kern-.125emX}}

\newcommand{\myset}[1]{\mathcal{#1}}

\newcommand{\myvar}[1]{\bm{#1}}

\DeclareMathOperator*{\NN}{NN}

\newcommand\munderbar[1]{%
  \underaccent{\bar}{#1}}

\begin{document}

\title{Differentiable Predictive Control for Large-Scale Urban Road Networks\\
\thanks{The information, data, or work presented herein was funded in part by the Advanced Research Projects Agency-Energy (ARPA-E), U.S. Department of Energy, under Award Number DE-AR0001565.  It may contain subject inventions with pending Intellectual Property at the United States Patent and Trademark Office.}
}
 
\author{
        Renukanandan Tumu \IEEEauthorrefmark{1}\IEEEauthorrefmark{2}, Wenceslao Shaw Cortez \IEEEauthorrefmark{1}, J\'an Drgo\v na \IEEEauthorrefmark{1}, \\ Draguna L. Vrabie \IEEEauthorrefmark{1},  Sonja Glavaski \IEEEauthorrefmark{1}
        \thanks{
            \IEEEauthorrefmark{1}
            \textit{Pacific Northwest National Laboratories},
            Richland, WA, USA 
            \{w.shawcortez, jan.drgona, draguna.vrabie, sonja.glavaski\}@pnnl.gov
        }
        \thanks{
            \IEEEauthorrefmark{2}
            \textit{University of Pennsylvania}
            Philadelphia, PA, USA 
            \{nandant\}@upenn.edu
        }
}

\maketitle

\begin{abstract}
Transportation is a major contributor to CO2 emissions, making it essential to optimize traffic networks to reduce energy-related emissions. This paper presents a novel approach to traffic network control using Differentiable Predictive Control (DPC), a physics-informed machine learning methodology.  We base our model on the Macroscopic Fundamental Diagram (MFD) and the Networked Macroscopic Fundamental Diagram (NMFD), offering a simplified representation of citywide traffic networks. Our approach ensures compliance with system constraints by construction. 
In empirical comparisons with existing state-of-the-art Model Predictive Control (MPC) methods, our approach demonstrates a 4 order of magnitude reduction in computation time and an up to 37\% improvement in traffic performance. Furthermore, we assess the robustness of our controller to scenario shifts and find that it adapts well to changes in traffic patterns. This work proposes more efficient traffic control methods, particularly in large-scale urban networks, and aims to mitigate emissions and alleviate congestion in the future.
\end{abstract}


\section{Introduction}
Transportation is the largest source of $\mathrm{CO_2}$ emissions in the United States, responsible for 38\% of energy-related emissions \cite{congressional_budget_office_emissions_2022}. Traffic networks in urban centers are notorious for traffic jams and delays, causing increases in travel time, and therefore, emissions. One way to reduce emissions is to control traffic in urban environments~\cite{papageorgiou_review_2003,NOAEEN2022116830,Qadri2020StateofartRO,HAN2023100104}.

A commonly used approach to model large-scale traffic networks for control is the Network Macroscopic Flow Diagram (NMFD)~\cite{JOHARI2021103334}. In this model, each region of a road network is grouped into a single unit, and the traffic flow through this region is modeled by a Macroscopic Flow Diagram (MFD). Two types of control have been applied to this system: Perimeter Control (PC) \cite{jiang_model_2021,DING2017300,FU201718} and Routing Guidance (RG) \cite{JOHARI2021103334,JIANG2024104440,Hou2022}. The PC approach can take the form of green light timing to red light timing ratios between adjacent regions. Jiang et. al. \cite{jiang_model_2021} use a partitioning approach to group regions together, and then solve for perimeter control using MPC. Fu et. al. \cite{FU201718} use stability analysis in a heirarchical approach to generate perimeter control, relying on an MPC approach to calculate the final control. The RG method can take the form of route planning software (such as Google or Bing Maps), or a global planning solution, that may be able to direct the flow of connected autonomous vehicles in the future. RG is more challenging to implement but can be impactful in alleviating congestion. Much of the existing state-of-the-art work uses Model Predictive Control (MPC) to control the NMFD system using permutations of PC and RG \cite{jiang_model_2021, sirmatel_stabilization_2021, sirmatel_economic_2018, geroliminis_optimal_2013}. Still other approaches use reinforcement learning (RL) to control traffic systems. These approaches are often model free, meaning that they do not rely on a modeling approach like the NMFD model to generate control. Noaeen et. al. \cite{NOAEEN2022116830} review RL approaches to traffic signal control. Without this model, RL approaches may not have notions of routing guidance, guarantees of performance or constraint satisfaction, and may require large numbers of rollouts.
These methods are able to achieve lower overall vehicle travel time than baseline controllers. 


Unfortunately, the MPC methods suffer from long solve times and occasional infeasibilities. Furthermore, these approaches do not scale well to networks with a large number of regions. We utilize Differentiable Predictive Control (DPC)~\cite{Drgona2024}, which helps address these concerns with MPC. DPC is a machine learning (ML) method that aims to approximately solve a parametric optimal control problem, just as in classical explicit MPC~\cite{Alessio2009}. DPC does this by training neural networks (NNs) control policies offline by using stochastic gradient descent on MPC-like loss functions backpropagated through differentiable models of the system dynamics. The loss function used to train these NNs is effectively the cost function of the MPC computed over a finite horizon rollout of the system dynamic simulations. The advantage of the DPC methodology is that it allows to leveraging the scalability of existing ML software and hardware infrastructure via libraries such as PyTorch~\cite{Paszke2019}.

In this paper, we apply DPC to traffic control via the NMFD model. 
Specifically, our work solves the parametric version of the Economic MPC problem formulation proposed by Sirmatel et al. \cite{sirmatel_economic_2018}.
Our contributions are:

\begin{itemize}
    \item We show that our method provides a five order of magnitude improvement in solve time, and a 19\% - 37\% improvement in performance as compared to the state-of-the-art Economic MPC method.
    \item We demonstrate that our method is robust to model mismatch.
    \item We show that online application of our method scales well to large-scale urban road networks with many regions. 
\end{itemize}

\section{Related Work}
The related work spans two categories: the modeling of the traffic system and the solution approach. We will discuss related traffic modeling approaches used for control, and different solution approaches used to generate control inputs to these systems.

There exist several methods for controlling traffic, some of which are summarized below. These approaches can be broadly categorized based on the underlying system model they use and the approach. The Cell Transmission Model (CTM) \cite{daganzo_cell_1994} is a model that operates on the level of individual traffic lights and road links. Prior work includes centralized receding horizon control of these CTM systems \cite{bianchin_gramian-based_2020}, as well as distributed control \cite{grandinetti_distributed_2019, pham_distributed_2022}. The Network Macroscopic Fundamental Diagram (NMFD) model operates at a higher level of abstraction and has been used extensively for control \cite{ren_data_2020,sirmatel_nonlinear_2020,yildirimoglu_equilibrium_2015,sirmatel_economic_2018,sirmatel_stabilization_2021,jiang_model_2021}, although potentially called an NFD or MFD-based model. While the CTM can generate higher fidelity simulations, it has a large number of parameters to identify, and requires a larger set of control inputs. One advantage is that CTM control inputs have simple real-world analogs. The NMFD model on the other hand, requires fewer parameters and control inputs, but the control inputs may be harder to implement on a real-world system. 

A previously used approach to solving this traffic flow problem is by using MPC~\cite{sirmatel_stabilization_2021,jiang_model_2021}. Specifically, we adopt the Economic MPC formulation used in prior work \cite{sirmatel_economic_2018}. 
MPC represents a well-established methodology that can optimize complex dynamical systems while satisfying operational constraints. However, the major disadvantage of MPC approaches based on online optimization are increased computational cost in real-time deployment.

A wide array of Reinforcement Learning methods have also been applied to the problem of traffic control \cite{NOAEEN2022116830, Qadri2020StateofartRO, han_leveraging_2023}. RL approaches suffer from high sample and computational complexity. One of the core ways to alleviate this is to find compact, state and action representations. RL approaches are varied, and most of the work surveyed thus far controls traffic lights directly. None of the $160$ approaches in Noaeen et. al. \cite{NOAEEN2022116830} implement a combination of routing guidance and perimeter control. This is difficult due to the complex constraints at play, the continuous state and action space, and the need for large amounts of data. Zhou and Gayah \cite{zhou_model-free_2021} use model-free RL to control a two region MFD using only PC, and require 50-75 epochs to approach the performance of MPC. The systems we seek to control are much larger, and implement both PC and RG.

To address this limitation, 
Drgo\v na et al.~\cite{Drgona2024} have developed a DPC methodology to allow offline solutions to the parametric optimal control problems arising in MPC.
DPC has been successfully applied in a number of applications~\cite{king_koopman-based_2022,drgona_deep_2021}; however, as of now, it has not yet been deployed to tackle challenging traffic control problems.
 DPC belongs to a family of physics-informed learning algorithms \cite{raissi_physics-informed_2019}, which leverage knowledge of physics or dynamics in conjunction with learned parameters. Generating control inputs can often be thought of as a constrained optimization problem, for which there are many approaches. Some of these approaches have been surveyed in Kotary et. al. \cite{kotary_end--end_2021}. Physics-informed learning methods have also been used to predict vehicle motion \cite{tumu_physics_2023} and materials modeling \cite{liu_multi-fidelity_2019}.
Agrawal et al. \cite{agrawal_differentiable_2019} develop methods to backpropagate through convex optimization approaches, and others use neural networks in conjunction with MPC to learn control \cite{amos_differentiable_2018,east_infinite-horizon_2019}. DPC differs from these approaches due to its flexibility to be used in scenarios where models may not be easily available, and its ability to generate solutions for non-convex problems.
Moreover, our work is the first to extend DPC to a networked system.

\section{Background}
\subsection{Notation}
The rest of the paper will use the following conventions for notation. Regular thickness variables, like $x$, will be used to express properties of subcomponents of a larger variable, such as update rules for parts of a variable. Bolded variables like $\myvar{x}$ will be used to represent the fully assembled version of the variable, and used to describe the usage of the complete variable in the system. Time, when noted in parentheses, like $(t)$, is continuous, and discrete when denoted in subscript like $\myvar{x}_k$. $A\sim D$ means that the variable $A$ varies according to the distribution $D$. ``$\NN$" denotes a feedforward neural network, and $\odot$ represents elementwise multiplication. Finally ``$\textrm{ODESolve}$" represents an arbitrary ordinary differential equation solver, which can be RK4 or forward Euler. This operation is used to discretize continuous dynamics.

\subsection{Modeling Approach}
\label{sec:model-description}

\begin{figure}[h]
    \centering
    \includegraphics[width=0.75\linewidth]{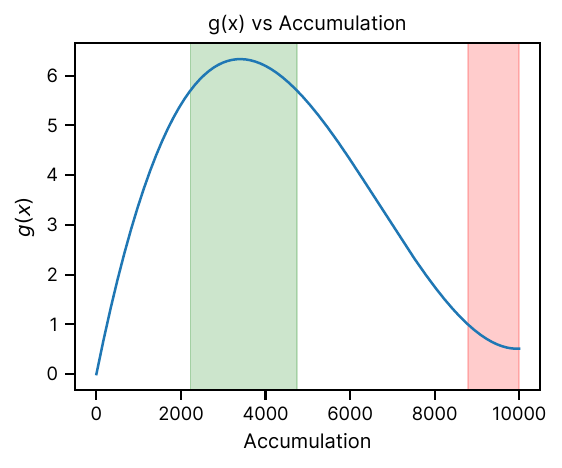}
    \caption{Macroscopic Flow Diagram. This figure has $[\max g(x),  \max g(x) \cdot 90\%$ highlighted in green, these are the optimal values of the MFD. The values of the MFD lower than $1$ with high accumulation are highlighted in red. This can be considered similar to a traffic jam.}
    \label{fig:mfd}
\end{figure}

Traffic flow in regions of cities can be described by the Macroscopic Fundamental Diagram (MFD), discovered in 1969 by Godfrey \cite{godfrey_mechanism_1969}. This diagram can be approximated by a third-order polynomial and describes the relationship between the amount of traffic flow and the amount of traffic inside a region. An example of this is shown in Figure \ref{fig:mfd}. This MFD modeling approach has been experimentally validated with data from Yokohama, Japan \cite{geroliminis_existence_2008}.

We chain several regions together in a Networked Macroscopic Flow Diagram (NMFD) Model, which is a collection of regions, each governed by its own MFD. We use this model as our abstraction for a citywide traffic network.

\subsection{Model Predictive Control}

When controlling our traffic system, we want to minimize the total vehicle time spent in the system, subject to control input constraints. These constraints include model-specific routing constraints that ensure the total volume of traffic leaving a region is preserved, to perimeter control constraints that ensure that traffic is not completely stopped. The MPC problem formulation in \Cref{eq:MPC-formulation} aims to solve the optimal control problem over the future horizon of control inputs. The advantage of an optimization-based MPC solution is that it can satisfy the state and input constraints while minimizing the cost function, which here is denoted by $\ell$. The state and input constraints are specified in \Cref{eq:mpc-formulation-sc,eq:mpc-formulation-ic}.
\begin{subequations}
    \small
    \label{eq:MPC-formulation}
    \begin{align}
    &\min_{u}                \quad \sum_{k=0}^{N} \ell(\myvar{x}_k) \\
    &\text{subject to: } \quad  \myvar{x}_0 = x_0 \\
    &\text{\quad $\forall k\in \{0, 1, \ldots, N-1\}$:}\\
                        &\qquad \myvar{x}_{k+1} = \text{ODESolve}( f( \myvar{x}_k,  \myvar{u}_k))) \\
                        &\qquad \myvar{x}_{k} \in \mathcal{X} \label{eq:mpc-formulation-sc}\\
                        &\qquad \myvar{u}_{k} \in \mathcal{U} \label{eq:mpc-formulation-ic}\\
    \end{align}
\end{subequations}

This optimization problem can be solved via solvers like IPOPT~\cite{Byrd99}. This problem is not convex, as the state update includes the MFD. This can make the solution of this problem difficult to find.

\subsection{Differentiable Predictive Control}

DPC is a method of solving parametric Optimal Control Problems (pOCP) \cite{Drgona2024,drgona_differentiable_2022}. This method solves a similar problem to MPC, but instead of minimizing over the control inputs themselves, the minimization is done over the policy weights $\beta$, using simulated rollouts of length $N$. Key components of DPC are the objective function $\ell$, the policy $\pi$, the system model $f$, and constraints that are at least once differentiable. The training data consists of initial states $x^i_0$, drawn from the known distribution $P_{x_0}$, and potentially additional sampled control parameters such as desired reference, constraints, or disturbance scenarios. DPC can be solved using stochastic gradient descent to optimize the parameters of control policy by minimizing an expectation of the control loss function that is estimated using $m$-number of sampled scenarios. The generic problem is presented in Equation \eqref{eq:DPC}. 
\begin{subequations}
\label{eq:DPC}
\begin{align}
&\underset{\beta}{\text{minimize}}     && \sum_{i=1}^m  \sum_{k=1}^{N-1} \ell( \myvar{x}^i_k)  \\
&\text{subject to}    && \myvar{x}^i_{k+1} =  \text{ODESolve}(f(\myvar{x}^i_k, \myvar{u}^i_k)) \\
&                     && \myvar{u}^i_k = \pi_\beta(\myvar{x}^i_k) \\
&                     && \myvar{x}^i_k \in \mathcal{X} \\
&                     && \myvar{u}^i_k \in \mathcal{U}  \\
&                     && \myvar{x}^i_0 \sim P_{x_0}
\end{align}
\end{subequations}

The advantage of DPC is that it can be implemented in modern ML frameworks such as PyTorch~\cite{Paszke2019} that allow us to leverage hardware accelerators such as GPUs.



\section{Methods}
Our objective is to control the NMFD system so that we minimize the total emissions, which we assume is related to the total vehicle time spent in the system. We aim to solve an economic optimal control problem originally posed in Sirmatel et al. \cite{sirmatel_economic_2018}. The system dynamics and the loss function are differentiable, making this system a good candidate for DPC. It is not immediately apparent that the constraints are differentiable. Still, we show that these constraints can be respected with careful construction of our control policies.
 One difference with MPC problem from~\cite{sirmatel_economic_2018} is that our DPC algorithm does not require a spawning matrix at run time, which is often difficult to estimate. MPC approaches require estimates of the system's future spawning rate.


\begin{figure*}[t]
    \centering
    \includegraphics[width=0.95\linewidth]{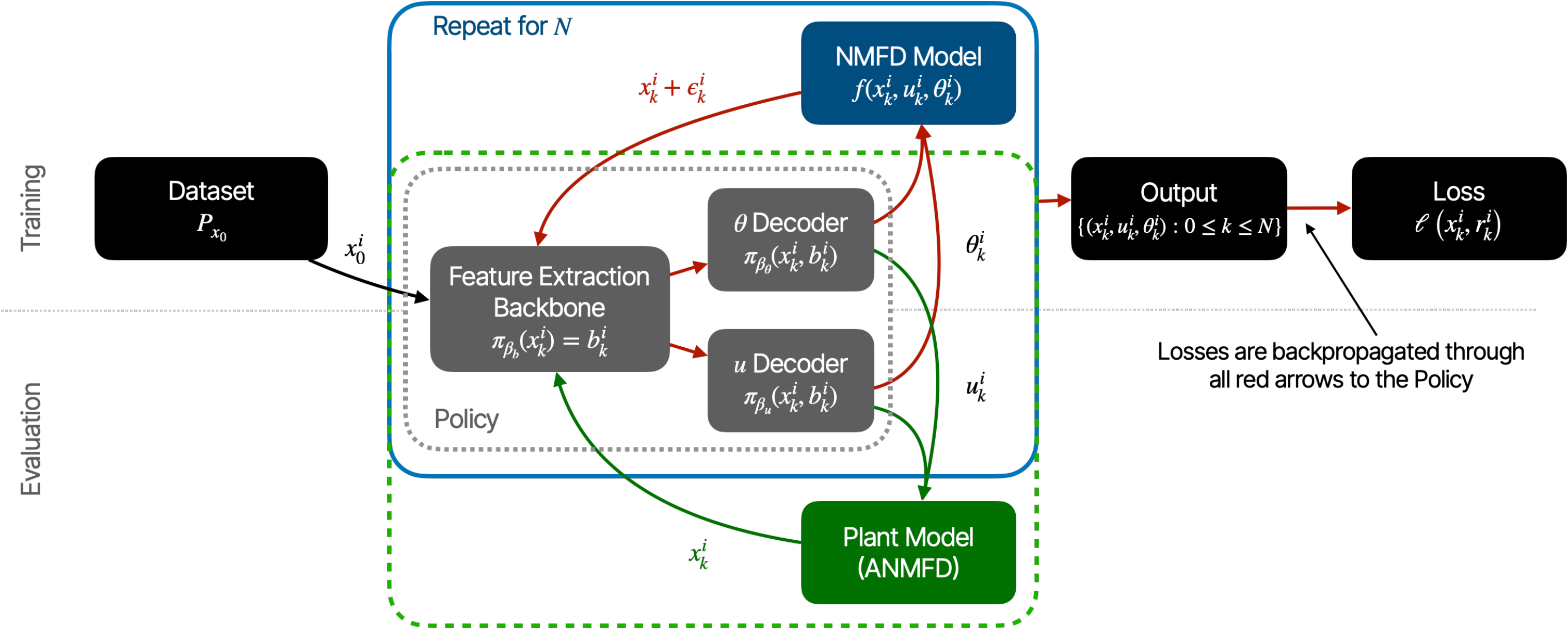}
    \caption{The figure shows the training and evaluation procedure. First, an initial condition is drawn from the dataset. This initial condition is provided to the model and policy. The policy generates a control input, which is provided to the model. The model generates a new state, noise is added, and this process is repeated for $N$ steps. At each step, the state and the control input are logged in the output. Our loss function is computed on the output and backpropagated to train the network. When evaluating, the policy and the plant model are evaluated in a closed-loop fashion.}
    \label{fig:training-diagram}
\end{figure*}

\subsection{Modeling}
\subsubsection{NMFD Model}
In this model, the traffic network is divided into $R$ regions, for which the set of regions is $\myset{R} = \{1,..., R\}$ and the set of adjacent regions to region $i$ is denoted $\myset{N}_i \subset \myset{R}$. Let $x_{ij} \in \mathbb{R}$, for $i,j \in \myset{R}$ denote the accumulation in region $i$ with destination $j$. These terms are considered the states of the system. Let $\myvar{x}$ be the concatenation of all $x_{ij}$ terms for all $i,j \in \myset{R}$. Each item must be non-negative, as specified in \eqref{eq:state-constraint}.
\begin{gather}
    \myvar{x} \in \myset{X} \iff x_{ij} \geq 0\quad \forall\, i,j\in\myset{R} \label{eq:state-constraint}
\end{gather}

For NMFD-based models, the dynamics depend on the MFD, a relationship defined in each region between the accumulation, i.e., the state, and the flow. To be more precise, we define the total accumulation of region $i$ with destination in all other regions as:
\begin{equation}
    x_i = \sum_{j\in \myset{R}} x_{ij}
\end{equation}

Let $g_i(x_i) \in \mathbb{R}_{\geq0}$ denote the `flow of region $i$' or the rate of vehicles exiting region $i$. The macroscopic fundamental diagram for region $i$ is usually determined by fitting a third-degree polynomial to a set of points determined empirically for region $i$ of the city. Approximations of the MFD are not limited to third-degree polynomials, but this is the convention observed in similar work \cite{sirmatel_economic_2018, sirmatel_stabilization_2021}. 
\begin{equation}\label{eq:MFD}
    g_i(x_i) = a_ix_i^3 + b_i x_i^2 + c_i x_i
\end{equation}

In addition to $g_i$, more terms affect the accumulation of vehicles in a given region. Vehicles will travel based on where they are and where their destination is. This is defined as follows:
\begin{equation}\label{eq:route flow 1}
    m_{ihj} = \theta_{ihj} \frac{x_{ij}}{x_i}g_i(x_i)
\end{equation}
\begin{equation}\label{eq:route flow 2}
    m_{ii} = \frac{x_{ii}}{x_i}g_i(x_i)
\end{equation}
where $\theta_{ihj} \in [0,1]$ is the routing term expressing vehicles exiting region $i$ that must transfer through region $h$ to get to destination region $j$. $\theta_{ihj}$ could be considered a control input if vehicles can be provided routing guidance. $\theta_{ihj}$ is the ratio of traffic in region $i$ with destination $j$ that proceeds to region $h$. Because these values represent ratios, $\sum \theta_{ihj} = 1$ and $0\leq \theta_{ihj} \leq 1$. We denote the fully assembled routing guidance control input as $\myvar{\theta}$. These constraints on $\myvar{\theta}$ are shown in Equation \eqref{eq:theta-constraints}.

The NMFD model also accounts for vehicles that will appear in regions as, for example, people reach their destinations or start their trips and add to the city's accumulation. Let $d_{ij}(k) \in\mathbb{R}$ denote the rate of vehicles  appearing in region $i$ with destination $j$ for a given time $k \in \mathbb{R}_{\geq 0}$. We refer to this as the spawn matrix.

The perimeter control input $\myvar{u}$, composed of elements $u_{i,j}\,\forall\, i,j \in \myset{R}$ of the system is the rate of transfer at the perimeter between regions. Let $u_{i,j} \in [\munderbar{u}, \bar{u}] \subset \mathbb{R}_{\geq 0}$ be the rate of transfer of vehicles between adjacent regions $i$ and $h$. From a microscopic perspective, this control input is treated as the ratio of green times at the intersections on the perimeter of each region to dictate how much flow to allow through the perimeters. The full set of control-related constraints are defined as follows:
\begin{gather}
    \myvar{u} \in \myset{U} \iff u_{ij} \in [\munderbar{u}, \bar{u}] \forall\, i\in\myset{R}, j\in \myset{N}_i  \label{eq:u-constraints}\\
    \begin{split}
        \myvar{\theta} \in \Theta \iff \left(\theta_{ihj} = 0\quad \forall\, i,j \in \myset{R}, h \notin \myset{N}_i \, \land \right.\\
        \left. \sum_{h=1}^n \theta_{ihj} =1 \quad \forall\, i,j \in \myset{R}\, \land\right. \\
        \left. 0 \leq \theta_{ihj} \leq 1 \quad \forall\, i,h,j \in \myset{R} \right) \label{eq:theta-constraints}
    \end{split}
\end{gather}

The dynamical model based on NMFD is written as follows:
\begin{subequations}\label{eq:nmfd-update}
\begin{align}
    \dot{x}_{ii}(t) &= d_{ii}(t) - m_{ii}(\myvar{x}) + \sum_{h \in \myset{N}_i} u_{hi} m_{hii}(\myvar{x}) \\
    \dot{x}_{ij}(t) &= d_{ij}(t) -\sum_{h \in \myset{N}_i} u_{ih} m_{ihj}(\myvar{x})  + \sum_{h \in \myset{N}_i, h\neq j} u_{hi} m_{hij}(\myvar{x})
\end{align}
\end{subequations} 

We can state the dynamics of the model as a function $\dot{\myvar{x}} = f(\myvar{x}, \myvar{u}, \myvar{\theta})$, which is a summarization of the model presented in Equation \eqref{eq:nmfd-update}. We discretize according to the equation in \eqref{eq:compact_ODE}, where the subscript $k$ refers to the system state at time $k$.

\begin{equation}
\label{eq:compact_ODE}
     \myvar{x}_{k+1} = \text{ODESolve}( f( \myvar{x}_k,  \myvar{u}_k, \myvar{\theta}_k))
\end{equation}

\subsubsection{Acyclic Plant Model}
The plant model used in the experiments is a variant of the NMFD model previously presented that prohibits traffic from traveling in graph cycles of length two. This model, the Acyclic NMFD (ANMFD) model, was used for experimental evaluations in Sirmatel et. al. \cite{sirmatel_economic_2018}. We reproduce the model equations here and expand upon the definition of $\theta$. We use the variable names $o$ and $g$ to denote the origin and previous region. $i$ denotes the current region, while $j$ and $h$ denote the destination and immediate next region respectively. The equation for the derivative of the state is given by \eqref{eq:ANMFD-dynamics}, as a function of the current time $k$.

\begin{subequations}
    \label{eq:ANMFD-dynamics}
    \begin{align}
        \begin{split}
            \dot{{x}}_{iiii}(t) &= d_{ii}(t) - m_{iiiii}, \qquad  \forall i \in \mathcal{R}
        \end{split}
        \\
        \begin{split}
            \dot{{x}}_{iiij}(t)  &= d_{ij}(t) - \sum_{h \in \mathcal{N}_i} U_{ih_k}m_{iiihj_k}, \\
                            &\qquad\qquad \forall i,j \in \mathcal{R},\quad j\neq i 
        \end{split}
        \\
        \begin{split}
            \dot{{x}}_{ogii}(t)  &= \sum_{f \in \mathcal{N}^*_g\setminus \{i\} } u_{gi_k}m_{ofgii_k} \\
                            &\qquad -m_{ogiii_k}, \\
                            &\qquad\qquad \forall o,g,i \in \mathcal{R},\quad g \in \mathcal{N}_{i},\quad o \neq i
        \end{split}
        \\
        \begin{split}
            \dot{{x}}_{ogij}(t)  &= \sum_{f \in \mathcal{N}^*_g\setminus \{i,j\}} u_{gi_k}m_{ofgij_k} - \\
                            &\qquad \sum_{h \in \mathcal{N}_i\setminus \{o,g\}} m_{ogihj_k}, \\
                            &\qquad\qquad \forall o,g,i,j \in \mathcal{R}, g \in \mathcal{N}_{i}, \\
                            &\qquad\qquad o \neq i,\quad o \neq j,\quad g\neq j,\quad j\neq i 
        \end{split}
    \end{align}
\end{subequations}

The Routing Guidance term, $\myvar{\theta}$ is also expanded to fit this new case.
\begin{subequations}
    \label{eq:ANMFD-theta}
    \begin{align}
        \theta_{ogihj_k}  &= \begin{cases}
            \theta_{ihg_k} & \textrm{if } h \neq g \\
            0               & \textrm{otherwise}
        \end{cases} \\
                        &\qquad \forall o,g,i,j \in \mathcal{R} \\
                        &\qquad g \in \mathcal{N}_{i}, h \in \mathcal{N}_{i}\setminus\{o\} \\
                        &\qquad o \neq i,\quad o \neq j,\quad g\neq j,\quad j\neq i \\
        \theta_{ogiii_k} &= 1.0 \\
        \theta_{iiihj_k} &= \theta_{ihj_k} \\
    \end{align}
\end{subequations}

The flow terms are defined by Equation \eqref{eq:ANMFD-m}. To translate these states back into the original system, we use the rule in Equation \eqref{eq:ANMFD-translation}.
\begin{gather}
    \label{eq:ANMFD-m}
    m_{ogihj_k} = \theta_{ogihj_k} \frac{x_{ogij_k}}{x_{i_k}}g(x_{i_k}) \\
    \label{eq:ANMFD-translation}
    x_{ij_k} = \sum_{o\in\mathcal{R}\setminus\{j\}} \sum_{g\in\mathcal{R}\setminus\{j\}} x_{ogij_k}
\end{gather}

\subsection{Problem Formulation}
We use the model as presented in Section \ref{sec:model-description}. Our total simulation runs for $T$ timesteps. The initial condition is denoted $X_0$, an $R\times R$ matrix of non-negative real values. We use the objective function we presented above, shown in Equation \eqref{eq:opt-obj}. We constrain the states to follow the dynamics of the system in \eqref{eq:state-transition}, and ensure that those states are positive in \eqref{eq:state-constraint}. It is not possible to have negative vehicles in the system. We ensure that the control inputs $\myvar{u}$ and $\myvar{\theta}$ obey their constraints in \eqref{eq:u-compliance}, that $\theta_{ihj}$ obeys system constraints in \eqref{eq:theta-compliance}. These constraints come directly from the model description in Section \ref{sec:model-description}, and the conditions in \eqref{eq:u-constraints} and \eqref{eq:theta-constraints} respectively. We also assume i.i.d. environmental noise $\epsilon_i^k$, drawn from a normal distribution with standard deviation $0.25$.

\begin{subequations}
\label{eq:prob-formulation}
    \begin{align}
        &\underset{\beta_{b},\beta_{\theta},\beta_{u}}{\text{minimize}}     && \sum_{i=1}^m  \sum_{k=1}^{N-1} \left\| \myvar{x}^i_k\right\|_1  \label{eq:opt-obj}\\
        &\text{subject to}    && \myvar{x}^i_{k+1} =  \text{ODESolve}(f(\myvar{x}^i_k, \myvar{u}^i_k, \theta^i_k)) + \epsilon^i_k \label{eq:state-transition} \\
         &                     && 
        \myvar{b}^i_k = \pi_{\beta_b}(\myvar{x}^i_k) \\
        &                     && 
        \myvar{u}^i_k = \pi_{\beta_u}(\myvar{x}^i_k, \myvar{b}^i_k) \\
        &                     && \myvar{\theta}^i_k = \pi_{\beta_\theta}(\myvar{x}^i_k, \myvar{b}^i_k) \label{eq:theta-gen}\\
        &                     && \myvar{x}^i_k \in \mathcal{X} \\
        &                     && \myvar{u}^i_k \in \mathcal{U} \label{eq:u-compliance}\\
        &                     && \myvar{\theta}^i_k \in \Theta \label{eq:theta-compliance}\\
        &                     && \epsilon^i_k \sim \mathcal{N}(0,\sigma_{\text{obs}}) \\
        &                     && \myvar{x}^i_0 \sim P_{x_0} \label{eq:dataset}
    \end{align}
\end{subequations}

The problem posed in Equation \eqref{eq:prob-formulation} is the Perimeter Control and Routing Guidance (PCRG) formulation of the problem. In cases where we are only solving the Perimeter Control (PC) version of the problem, the lines \eqref{eq:theta-gen} and \eqref{eq:theta-compliance} are removed, and $\myvar{\theta}$ is set to a default value for all timesteps. The default value is calculated based on Djikstra's shortest path algorithm \cite{dijkstra1959note}. The resulting architecture presented in Figure \ref{fig:training-diagram} is implemented using the NeuroMANCER scientific machine learning library~\cite{Neuromancer2023}.


Recalling that our objective is to reduce the overall emissions of vehicles in the network. We assume that the minimization of travel time for a single vehicle minimizes the emissions, as the vehicle will be running for less time.
The sum of the state for each time step in the simulation multiplied by the size of the timestep $dt$ gives the total time spent. This expression: $\sum_{k=0}^T \left\|x_k\right\|_1 dt$, when minimized, is the same as \eqref{eq:opt-obj}, which we use as our objective function. This is the same objective used in the Economic MPC approach \cite{sirmatel_economic_2018}.

\subsection{Policy Network}
We use a feature extraction backbone consisting of a Multi-Layer Perceptron (MLP) neural network paired with a SoftExponential non-linearity that takes the system state $\myvar{x}$ as features. This feature extractor is followed by two control-decoder networks, one to generate control inputs $\myvar{u}$ and another to generate $\myvar{\theta}$. These networks are designed to satisfy the system constraints as described in \Cref{eq:u-constraints,eq:theta-constraints}. Without satisfaction of these constraints, the control inputs will not be suitable for application to the system.

\subsubsection{Feature Extraction Backbone}
This neural network, $\pi_{\beta_b}(\myvar{x})$, takes the last state as input and produces a $n_{\xi}-\text{dimensional}$ feature vector $\myvar{b}$, and is parameterized by the weights of a neural network, denoted $\beta_b$. This feature vector will be used by the control input decoders.

\subsubsection{Perimeter Control Decoder Network}
The decoder network for $\myvar{u}$: $\pi_{\beta_u}(\myvar{x}, \myvar{b})$, and produces the control input $\myvar{u}$. The constraints for this control input are described in \eqref{eq:u-constraints}. The policy is composed of a neural network parameterized by $\beta_u$, and an activation function which is shown below. The function $\sigma$ is the sigmoid activation function $\sigma(\nu) = 1/(1+\exp(-\nu))$.
\begin{equation}
    \pi_{\beta_u}(\myvar{x}, \myvar{b}) = \sigma\left(\NN_{\beta_u}\left(\myvar{b}\right)\right)\cdot(\bar{u}-\munderbar{u}) + \munderbar{u}
\end{equation}
This function ensures the output of the PC decoder network is between $0$ and $1$, then scales this to the interval $[\munderbar{u},\bar{u}]$.

\subsubsection{Routing Guidance Decoder Network}
The decoder network for $\myvar{\theta}$: $\pi_{\beta_\theta}$, uses $\myvar{b}$, and produces the control input $\myvar{\theta}$. We are also given an adjacency matrix for the graph, $A$, which takes value $1$ if two regions are adjacent, and $0$ otherwise. The constraints that apply to $\myvar{\theta}$ are shown in Equation \eqref{eq:theta-constraints}.
These constraints ensure that the routing terms add up to $1$ for each origin and destination pair, and that traffic is only routed to adjacent regions. In order to ensure these constraints are obeyed by construction, we first apply a mask over the output of the neural network to ensure the adjacency constraint is not violated. We do this by multiplying by a mask $\alpha$, element-wise with $\myvar{\theta}$. The mask is defined as $\alpha_{ihj} = A_{ih}\quad \forall i,h,j \in \myset{R}$. This drives all elements of $\myvar{\theta}$ that correspond to non-adjacent regions to $0$, which ensures that our adjacency constraint is met. Then, we apply a softmax over only the non-zero elements of the $h$ dimension of the masked output matrix, which ensures the satisfaction of our other constraints, namely that the routing ratios for any pair $i,j$ sum to $1$. 
The resulting policy network to generate the control input $\myvar{\theta}$ is
\begin{align}
    \pi_{\beta_\theta}(\myvar{x},\myvar{b}) &= \text{softmax}_{h\in\mathcal{N}_i}\left(\alpha \odot \NN_{\beta_\theta}\left(\myvar{b}\right)\right)
\end{align}



\subsection{Training}
Our training process is summarized in Figure \ref{fig:training-diagram}.
We train in batches, where each batch contains a number of initial conditions sampled from $P_{x_0}$. We take the initial conditions from the dataset for each batch and provide them to the policy and the model. The policy and the model are evaluated sequentially for $N$ iterations. Finally, our loss is computed on these outputs and is backpropagated to our policy. We use the Adam optimizer with weight decay to update the weights of our policy. Our DPC controller was trained with a learning rate of $1\cdot 10^{-4}$, a weight decay factor of $1\cdot 10^{-6}$, and a batch size of $256$. The size of the feature vector $n_\xi = 128$, as were all of the hidden sizes in the network. 

When the policy with Routing Guidance and Perimeter Control is trained, it is trained in an alternating fashion, where the weights for the $\myvar{\theta}$ Decoder network are frozen while the weights for the $\myvar{u}$ Decoder are updated, and vice versa. In our experiments, we alternated which network we trained every epoch. Implementing this switching methodology improves the performance of the DPC PCRG approach significantly, producing a $\sim 48\%$ improvement in total accumulation.

While the $N$ acts as a kind of planning horizon, it is important to note that the same neural network accumulates all of the gradients from states all throughout the scenario, and is therefore biased to act in ways that may be more globally optimal.

\section{Results and Discussion}
We ran three experiments to investigate the performance of the proposed DPC approach. First, we evaluated the performance of our approach on a benchmark scenario. Second, we investigated the robustness of the trained DPC policy to model mismatch, where the scenario the policy was evaluated on differed from the training policy. Finally, we evaluated the scalability of our approach without hardware acceleration. We compared our approach to three baselines: a no-control benchmark, and the MPC approach with and without routing guidance from \cite{sirmatel_economic_2018}. In the no-control approach, we set all components of the control input to be equal to $\bar{u}$, and the routing term to be equal to the nominal measured value $\theta$. The MPC approach used below has a planning horizon of $8$ timesteps, greater than the horizon of $7$ presented in \cite{sirmatel_economic_2018} past which there are diminishing returns. The types of control we used were Perimeter Control (PC) and Perimeter Control and Routing Guidance (PCRG). All approaches were trained on the NMFD model, not the plant model.

\subsection{Economic MPC Comparison}
\begin{table*}[t!]
\centering
\sisetup{
    detect-weight=true, detect-inline-weight=math, 
    table-format=+1.1, table-auto-round, 
    scientific-notation = true, exponent-mode= scientific,
    round-mode = figures, round-precision = 3
}%
\label{tab:eco-mpc-results}
\caption{Overall Accumulation Results\\
PC=Perimeter Control,PCRG=Perimeter Control and Routing Guidance
}
\begin{tabular}{l|S[table-format=9.2]|S[table-format=8.2]|S[table-format=5.3]|S[table-format=5.3]}
\textbf{Algorithm} & \shortstack{\textbf{Total}\\ \textbf{Accumulation}\\($\text{veh} \cdot \text{s}$) $\downarrow$} & \shortstack{\textbf{Final}\\ \textbf{Accumulation}\\($\text{veh}$) $\downarrow$}  & \shortstack{\textbf{No-Control}\\\textbf{Improvement}\\($\text{veh} \cdot \text{s}$) $\uparrow$} & \shortstack{\textbf{Total}\\ \textbf{Evaluation} \\\textbf{Time} (s) $\downarrow$}  \\
\hline
No-Control  & \num{177272484.23339346}      & \num{25261.575068223898}  &                                  &  \\ \hline

MPC PC      & \num{169233838.89815277}      & \num{24230.061868118028}  & \num{8038645.33524069}                       & \num{315.4923720359802}  \\

DPC PC      & \bfseries \num{111586785.0}   & \bfseries \num{1567.731531297788}  & \bfseries \num{65685699.23339346}    & \bfseries \num{0.586}  \\ \hline

MPC PCRG    & \bfseries \num{91738273.56483985}       & \num{9201.871822133788}  & \num{85534210.66855361}                      & \num{3729.9903190135956}  \\

DPC PCRG    & \num{96069397.5}    & \bfseries \num{844.1873367467701}  & \bfseries \num{81203086.73339346}   & \bfseries \num{0.7558653354644775}  \\
\end{tabular}
\end{table*}
\begin{figure}[h]
    \centering
    \includegraphics[width=0.75\linewidth]{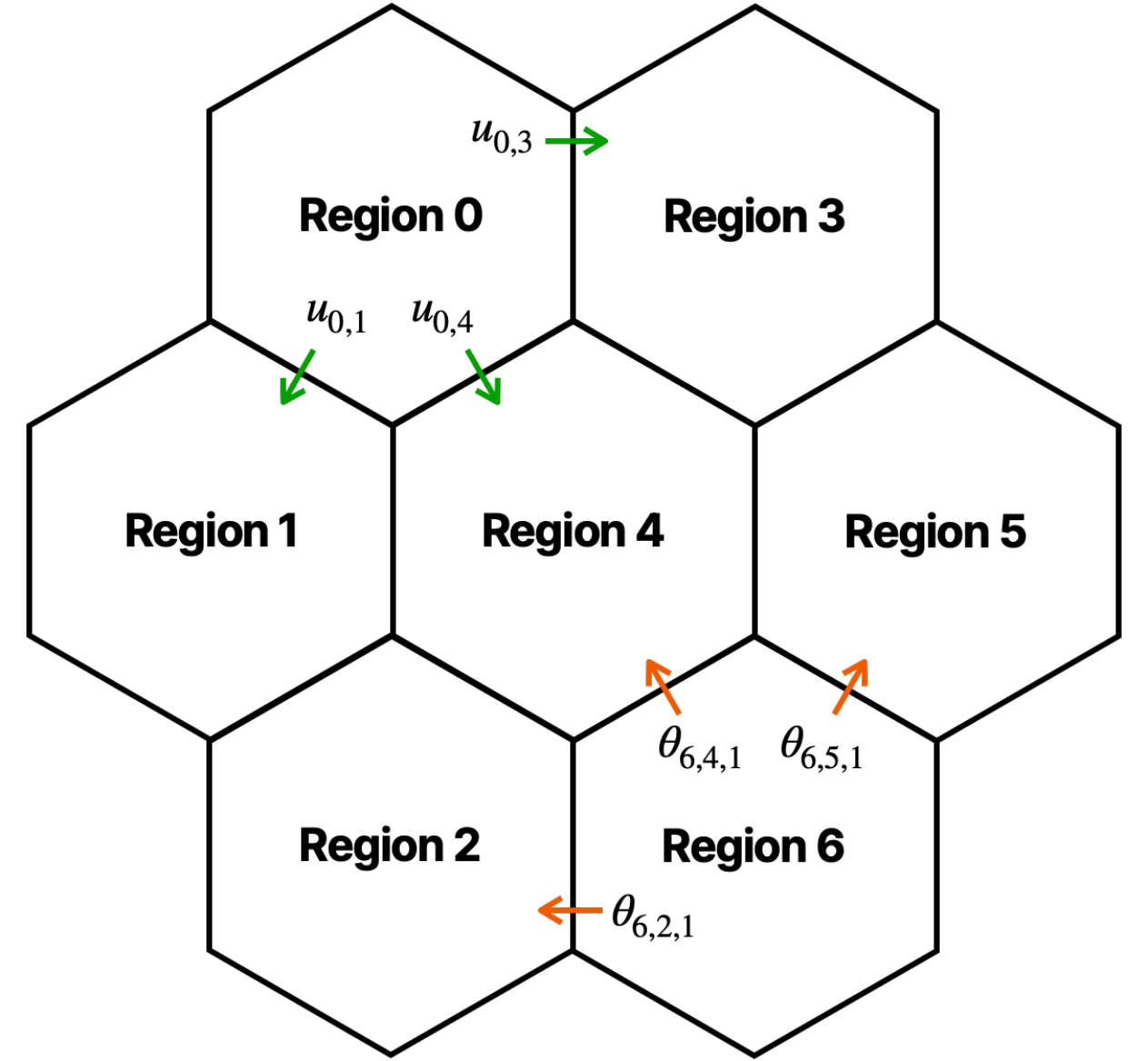}
    \caption{This figure shows the layout of the regions in the Economic MPC Scenario. There are seven regions here, and only adjacent regions are connected in the NMFD model.}
    \label{fig:eco-scen-regions}
\end{figure}

\begin{figure}
    \centering
    \includegraphics[width=\linewidth]{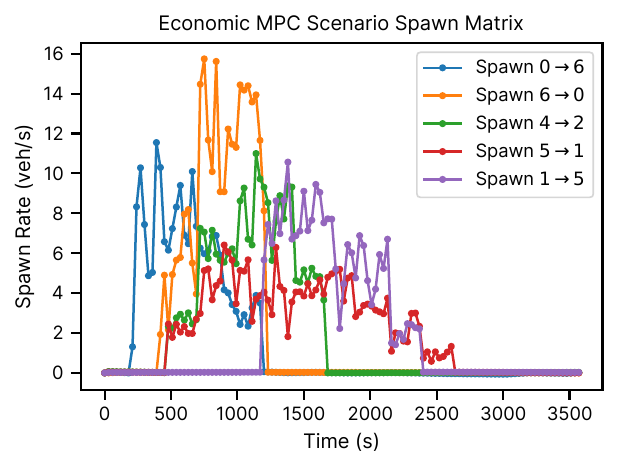}
    \caption{This figure shows the spawn matrix for the Economic MPC Scenario. These spawning rates for traffic describe the scenario's evolution over time.}
    \label{fig:spawn-matrix}
\end{figure}

To compare the performance of our approach to the benchmark approach, we adopt the same scenario as in \cite{sirmatel_economic_2018}. The network topology is shown in Figure \ref{fig:eco-scen-regions}. Our MFD uses the parameters $a=4.133 \cdot 10^{-11}, b=-8.282 \cdot 10^{-7}, c=0.0042$. The total experiment length $T=240$, and the timestep $dt=30$. The horizon $N$ for DPC was set to $240$, the length of the total experiment. We apply a measurement normally distributed measurement noise $\epsilon \sim \mathcal{N}(0.0, 0.25)$. All policies are provided a noisy version of the system state. The control limits take values $\bar{u}=0.9$ and $\munderbar{u}=0.1$. We generate the routing guidance term $\theta$ by calculating the shortest paths between each origin-destination pair. The initial state has no vehicles, and the spawn matrix controls the scenario evolution. The scenario specification matches that used in \cite{sirmatel_economic_2018}. The values of the spawn matrix we used are shown in Figure \ref{fig:spawn-matrix}. Traffic in this scenario flows between one of three pairs of nodes: between Region 0 and 6, between Region 5 and 1, and from Region 4 to Region 2.

\begin{figure}[h]
    \centering
    \includegraphics[width=0.9\linewidth]{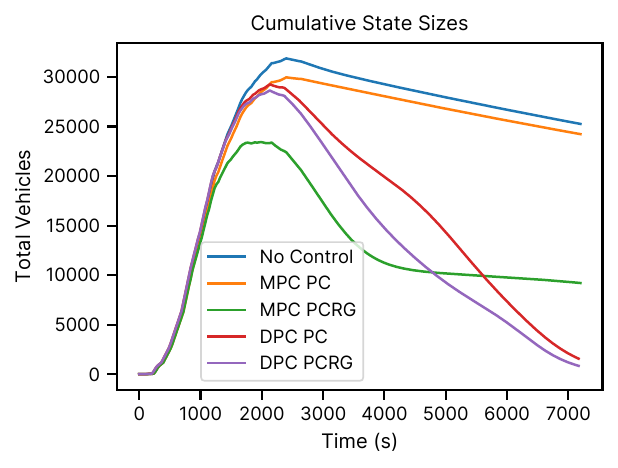}
    \caption{This figure shows the total accumulation of traffic in the system at each timestep. In the PC setting we can see that the DPC achieves lower peak accumulation, and lower overall accumulation. In the PCRG setting, we see that the MPC achieves lower peak accumulation, but fails to dissipate the traffic by the end of the scenario.}
    \label{fig:eco-total-accumulation}
\end{figure}

The total accumulation over time is shown in Figure \ref{fig:eco-total-accumulation}. This plot shows that the performance of the DPC approach, with and without routing guidance control, outperforms the MPC and No Control approaches. We can also see that the performance of both DPC approaches is largely similar. The MPC with routing guidance control controls the system to achieve lower peak accumulation levels but has a higher final accumulation than the DPC approaches. The final accumulation totals are presented in Table \ref{tab:eco-mpc-results}. We can see that the DPC approach with only Perimeter Control achieves a $33.72\%$ improvement over the MPC approach in total vehicle seconds. The DPC with Routing Guidance and Perimeter Control accumulates $4.79\%$ more vehicle seconds traveled over the total scenario duration. While the total accumulation is higher for DPC, the traffic left in the system at the end of the scenario is $90\%$ less. The DPC approach also keeps all of the regions from reaching a jam state, while the MPC induces a traffic jam.

\begin{figure*}[t]
    \centering
    \includegraphics[width=\linewidth]{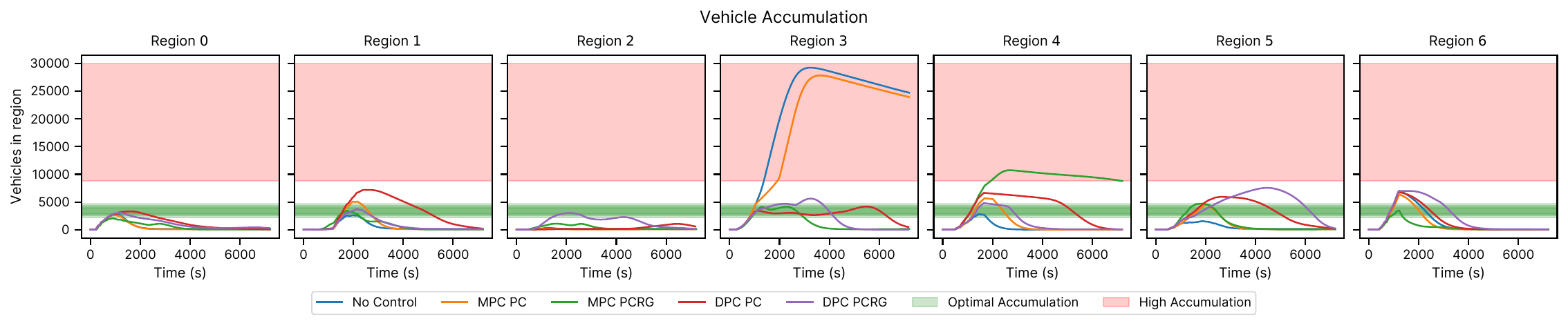}
    \caption{This figure shows the accumulation in each of the seven regions over time. PC=Perimeter Control, RG=Routing Guidance. We can see the comparison of traffic accumulation in each region for each method. The optimal accumulation shading represents the accumulation in each node that will produce the optimal values of the MFD, within 10\%. It is best for the accumulation of traffic to be kept in this range. The red range represents areas of suboptimal accumulation, where the values of the MFD are less than $1$. We can see that the DPC approaches keep the traffic in the green areas for longer than their MPC counterparts. The accumulation of the PCRG MPC in Figure \ref{fig:eco-total-accumulation} is largely in Region 4, and dissipates poorly due to being in the red-zone of the MFD.}
    \label{fig:eco-region-accumulation}
\end{figure*}

In observing the region-wise accumulation in Figure \ref{fig:eco-region-accumulation}, we can see the difference in the accumulation patterns of the tested algorithms. In this plot, the green zone represents the node accumulation that achieves the highest flow, as determined by the MFD $g(x_i)$ in Figure \ref{fig:mfd}. The green zone shows the range of values that are within $10\%$ of the maximum value of $g(x_i)$. The red zone represents high accumulations, which provide MFD values of less than 1, leading to slowed vehicle flow.

Focusing on Region 3, we can see that the best-performing approaches (DPC with and without region guidance, and MPC with region guidance) maintain the accumulation in the green zone for the longest. This follows intuition, as Region 3 is central to the system, and jams in this region could result in a network-wide traffic slowdown, especially when Routing Guidance is not provided. However, the MPC approach with Routing Guidance reaches high levels of accumulation in Region 4, resulting in a jam that takes a long time to dissipate.

Despite having access to the same control mechanism, the DPC PC approach performs much better than its MPC counterpart. This seems to be because the MPC, with its' shorter prediction horizon, is unable to keep traffic from accumulating in Region 3. This stems from the computation resources required to compute MPC solutions. We believe that the reason DPC is able to improve upon this is its' training process, which incorporates information from the entirety of the scenario to determine updates to the policy, providing a degree of foresight that is not present in the MPC. This may make the DPC approach more able to make short-term sacrifices for long-term benefits in the cost function.

\subsection{Robustness to Noise}
We also aim to show that our approach can tolerate shifts in the behavior of the underlying system by perturbing the spawn matrix.  We start with the vectors describing the spawn rates for traffic in our system over time, shown in Figure \ref{fig:spawn-matrix}, and add normally distributed noise with a mean of $0$ and a standard deviation of $\sigma_{\text{scenario}}$. We then clip the resulting vector, so that the spawn rates are always non-negative. This is intended to show the performance of our DPC approach in cases where our original estimation of the spawn rates is incorrect. We sample $20$ scenarios with $\sigma_{\text{scenario}} \in [0.0, 0.25, 0.5, 0.75, 1.0, 1.5, 2.0, 3.0, 5.0]$.

It is of note that while the overall traffic accumulation is high, the values of these spawn rates are fairly low, with a maximum value of around $16$. Therefore, while the noise standard deviation may seem low, a standard deviation for noise ranging from $0$ to $5$ does provide a reasonable spread of estimation error. To control for the effects of lower spawn rates, we measure the performance of DPC against a No-Control baseline lowereach noised scenario. A lower improvement represents a degradation in performance.

\begin{figure}[h]
    \centering
    \includegraphics[width=\linewidth]{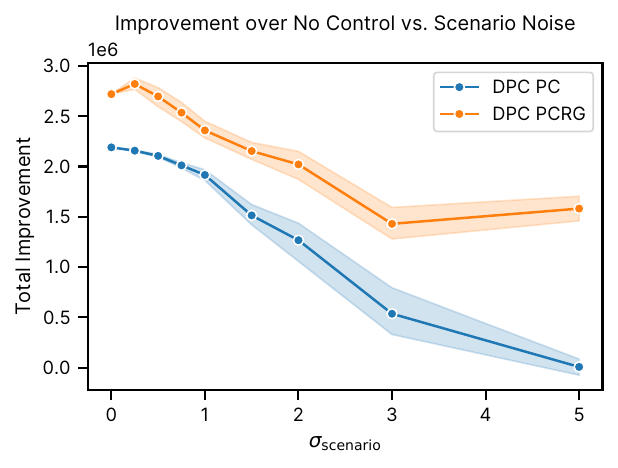}
    \caption{This figure shows the performance of the DPC controller with increasing levels of noise in the spawn matrix $d$. The shaded region represents the 95\% confidence interval of the test improvement. The improvement is presented with respect to a no-control approach, in order to account for baseline scenario difficulty.}
    \label{fig:noise-tolerance}
\end{figure}

The results of this experiment are reported in Figure \ref{fig:noise-tolerance}. We can see that the performance of the DPC algorithms degrades as the scenario changes. The performance of the Perimeter Control approach degrades slower, producing similar performance while the standard deviation is less than $1$, after that, the performance degrades quickly. The Perimeter Control and Routing Guidance approach experiences a much faster dropoff, indicating that the method is much less able to tolerate a shifted scenario. While the performance is not ideal in high-noise environments, the approach is comparable to the no-control approach.

Overall, this experiment shows that the degradation of performance for DPC approaches is graceful and tolerant of scenario shift. When deploying DPC approaches, it may be necessary to detect these scenario shifts, and apply a safe controller, or fall back to an MPC approach. We aim to tackle the issue of scenario shift in future work.

\subsection{Region Scaling Results}
\begin{figure}[h]
    \centering
    \includegraphics[width=\linewidth]{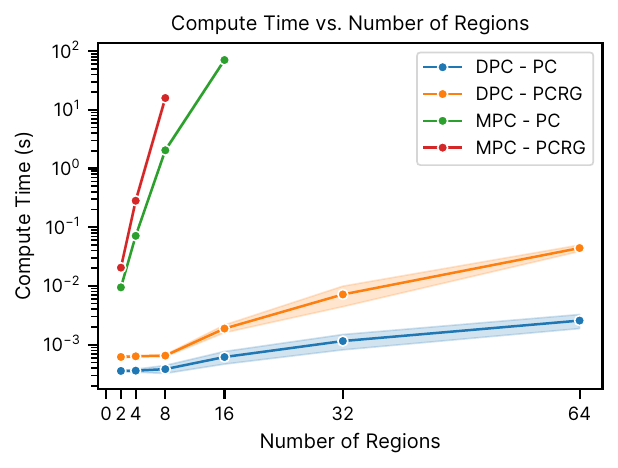}
    \caption{This figure shows the compute times of MPC and DPC with varying number of regions. The MPC controller failed to compute when the number of regions was greater than or equal to 32.}
    \label{fig:region-scaling}
\end{figure}

We also aim to show that our approach scales better to a greater number of regions. For each number of regions, we create a fully connected graph with initial states that take random values in the interval $[0, 100)$  with uniform probability. The MFD is also randomly selected. Due to the fully connected topology of the system, we expect that this system is quite easy to solve for the MPC. The results of this experiment should be interpreted as giving the order of compute time requirements for each type of system. In this experiment, we are only measuring the compute time required to produce a policy output. We use the CPU to compute the MPC and the DPC. Each system is evaluated for ten timesteps, and the time taken to produce an output is recorded. For the MPC evaluations, the time taken to set up the problem, as well as the first solve are not included, as they are much larger than subsequent solutions, which have been warm started. We evaluate the approaches where the number of regions $R \in [2^1, 2^2, 2^3, 2^4, 2^5, 2^6]$. Results were not reported for the MPC Perimeter Control approach where $R>16$ or the MPC PCRG approach where $R>8$, as the MPC would not compute on our system. The results of this experiment are reported in Figure \ref{fig:region-scaling}.

We can see that the compute times for the DPC approaches scale much better with the number of regions than the MPC. Further improvements are expected with the usage of a GPU, which should reduce the time taken for DPC evaluation even further. The time complexity scales with the size of the matrices that we are multiplying, which only results in an increase in compute time for the first and last layers. The MPC approach requires a problem setup with a number of constraints, and then requires a search to compute the optimal control. The experiments presented show that DPC approaches are more scalable in the number of regions than MPC.

\section{Conclusion}
We have presented a Differentiable Predictive Control (DPC) algorithm for large-scale urban road networks which produces better performance with lower evaluation cost than existing state-of-the-art approaches. In the Perimeter Control setting, we show a $34\%$ improvement in total vehicle seconds over the Model Predictive Control (MPC) approach, with five orders of magnitude improvement in solve time. In the Perimeter Control and Routing Guidance setting, our DPC approach produces a $90\%$ decrease in vehicles present at the end of the simulation with a similar reduction in solve time. Despite being trained on a single scenario, our approach generalizes to small scenario shifts well, with a smooth degradation in total vehicle seconds as the scenario shifts. Finally, we show that compared to MPC, our approach scales better to a large number of regions, thus better accommodating large-scale traffic networks.

There are a number of avenues for future work to explore, including data and model parallel implementations to make DPC run faster, improved training methods to allow for faster learning and finetuning, and the potential to deploy online variants of DPC that update the controller based on realized system outcomes.

\section{Acknowledgements}
The information, data, or work presented herein was funded in part by the Advanced Research Projects Agency-Energy (ARPA-E), U.S. Department of Energy, under the grant titled: Autonomous Intelligent Assistant (AutonomIA): Resilient and Energy-Efficient City-wide Transportation Operations.

\bibliographystyle{IEEEtran}
\bibliography{IEEEabrv,references,refs_jan}

\end{document}